\documentclass[a4paper]{article}

\usepackage{INTERSPEECH2019}

\title{Learning Shared Encoding Representation for End-to-End Speech Recognition Models}
\name{Thai-Son Nguyen, Sebastian St\"uker, Alex Waibel}
\address{Institute for Anthropomatics and Robotics, Karlsruhe Institute of Technology}
\email{thai.nguyen@kit.edu}

\begin{document}

\maketitle
\begin{abstract}
In this work, we learn a shared encoding representation for a multi-task neural network model optimized with connectionist temporal classification (CTC) and conventional framewise cross-entropy training criteria. Our experiments show that the multi-task training not only tackles the complexity of optimizing CTC models such as acoustic-to-word but also results in significant improvement compared to the plain-task training with an optimal setup. Furthermore, we propose to use the encoding representation learned by the multi-task network to initialize the encoder of attention-based models. Thereby, we train a deep attention-based end-to-end model with 10 long short-term memory (LSTM) layers of encoder which produces 12.2\% and 22.6\% word-error-rate on Switchboard and CallHome subsets of the Hub5 2000 evaluation.
\end{abstract}


\section{Introduction}

The recent trend of automatic speech recognition (ASR) research is to simplify the recognition process by using a single neural network to approximate the direct mapping from acoustic signals to textual transcription. Connectionist temporal classification (CTC)~\cite{graves2006connectionist,graves2014towards} is an appealing approach thanks to the ability to directly model the alignment between acoustic observations and label sequences. In end-to-end CTC systems~\cite{amodei2016deep,miao2015eesen,zweig2017advances}, the traditional components such as pronunciation lexicons, cluster trees, framewise alignment or Hidden Markov Model (HMM) topology are \textit{not} mandatory. The introduction of sequence-to-sequence attention-based model~\cite{chorowski2015attention,bahdanau2016end,chan2016listen} for speech recognition advances the simplification by learning both acoustic and language models with a single neural network.

The aforementioned works have shown the potential of the CTC framework to be able to jointly learn to predict the labels and to align between inputs and outputs. However, the disadvantage of CTC is training complexity, i.e. the convergence is not guaranteed or optimization gets stuck in local optima~\cite{sak2015learning,sak2015fast}. Data sparsity is also known to be a hindrance to train CTC models effectively, as can be seen from the work of~\cite{audhkhasi2017direct} reporting that without a pretraining initialization, a word-based CTC model (or acoustic-to-word) is much harder to converge on the well-known Switchboard training data set.

Our work is motivated by the findings in our previous study~\cite{Nguyen2018}, in which the correlation between the phone probabilities estimated by CTC models and the hard labels produced by traditional frame-wise alignment models was found. This evidence indicates the correlation between two training schemes: sequence-wise prediction with CTC and frame-wise classification typically using a cross-entropy (CE) optimization criterion. In this work, we propose a novel architecture in which a shared neural network is used to train acoustic models with both CTC and framewise CE criteria, that we treat with a multi-task learning perspective. Our experiments show that the proposed multi-task learning model not only tackles the complexity of CTC training but also improve the performance of both tasks on several setups. We then further propose to use the shared multi-task neural network to initialize the encoder of attention-based models. By doing so, we can easily train a deep sequence-to-sequence speech recognition model which produces attractive results on Switchboard benchmark.
 
\section{Multi-task Learning of CTC \& Framewise CE}
\label{sec:mlt}

In this section we review two popular optimization criteria CTC (Section \ref{ssec:ctc_task}) and framewise CE (Section \ref{ssec:ce_task}) frequently used in training neural network-based acoustic models. Then our proposed network architecture for combining these criteria is described in section \ref{ssec:join_learning}.

\vspace{-0.2cm}
\subsection{CTC Task}
\label{ssec:ctc_task}

Given an audio utterance~$x$~and the corresponding transcript $z$ (a sequence of labels), the CTC framework estimates alignments between the utterance and the transcript as a latent variable, dubbed the CTC path. Let us denote $y^\pi_t$ as the posterior probability that the neural network assigns to the corresponding label of $\pi$ at time $t$, then the CTC loss function is defined as the sum of the negative log likelihoods of the alignments for each utterance:

\vspace{-0.4cm}
\begin{eqnarray*}
	\mathcal{L}_{CTC} = - \sum \limits_{\pi} p(\pi|x) = - \sum \limits_{\pi} \prod \limits_{t} y^\pi_t \\ 
\end{eqnarray*}
\vspace{-0.8cm}

\noindent In order to optimize towards the CTC criterion, \cite{graves2006connectionist,graves2014towards} proposed to use the forward-backward algorithm which efficiently computes the gradient with respect to the activation of neural networks for every input frame.

\subsection{Framewise CE Task}
\label{ssec:ce_task}

Assuming that HMM-based speech recognition uses Viterbi forced alignment to obtain a \textit{ state} label (i.e., context-dependent phoneme) $l$ from the ground-truth transcript $w$ for each input frame of a training utterance $x$, a neural network model is trained by optimizing the CE loss function, to model this \textit{state} distribution:

\vspace{-0.4cm}
\begin{eqnarray*}
	\mathcal{L}_{CE} = \sum_{t=1}^{|x|} \sum \limits_{l} \delta(l,l_t) \textrm{ log} y^t_l\\ 
\end{eqnarray*}
\vspace{-0.8cm}

\noindent where $\delta(l,l_t)$ is the Kronecker delta and $y^t_l$ is the network output for the state $l$ at the frame $t$. According to the HMM assumptions, a model which minimizes the CE loss approximately maximizes the likelihood of the input $p(x|w)$.

\vspace{-0.2cm}
\subsection{Learning Shared Encoder}
\label{ssec:join_learning}
\vspace{-0.1cm}

The neural network models trained with CTC and conventional framewise CE criteria provide very different label distributions even when the used label sets are identical, because they learn the mapping function at different scales. However, the models resulting from optimizing either CTC or framewise CE criterion may eventually share similar traits in representation as potentially shown in \cite{Nguyen2018}. This observation motivates us to use these loss functions jointly. Here we consider each of them as a task and the aim of our work is to establish a shared underlying neural network model to efficiently learn both tasks in parallel. 

\begin{figure}[t]
	\centering
	\includegraphics[width=0.99\linewidth]{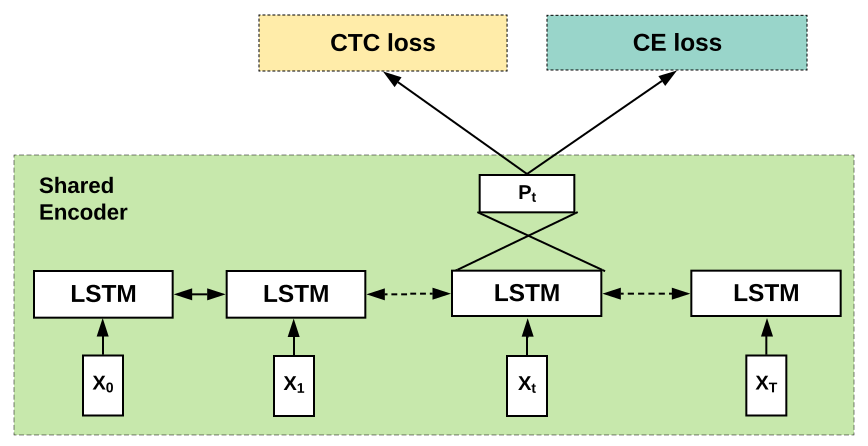}
	\vspace*{-0.1cm}
	\caption{Learning shared encoder with CTC and CE criteria.}
	\label{fig:multitask_network}
	\vspace*{-0.5cm}
\end{figure}

Figure \ref{fig:multitask_network} illustrates our proposed network architecture combining the two training criteria. Basically, the entire long short-term memory (LSTM) architecture used for encoding input sequences is shared, only two output layers are separated to perform the specific tasks. This structure allows gradients to be propagated back to the encoding network as early as possible. We further added a small projection layer into the shared network, on top of the LSTM layers. The projection layer was found to speed up training and improve convergence \cite{sak2014long}. In this case, it significantly reduces the number of parameters in the task-specific layers, so that it pushes the shared layers to learn their needed representation.

\vspace{-0.2cm}
\section{Optimizing Multi-task Training}
\label{sec:optimizations}

\subsection{Loss Combination}
CTC loss is typically computed on entire training utterances so that the model can effectively learn the prediction for the labels and their corresponding alignments at the same time. Different from that, the framewise CE loss does not consider each utterance in its entirety, but instead it is defined over individual samples of acoustic frames and state labels. In practice, several works~\cite{sak2014long,li2015constructing,zeyer2017comprehensive} have found dividing training utterances into subsequences of fixed-sized chunks (e.g., 50 frames) achieves optimal performance when modeling with LSTMs. To obtain optimal performance for both tasks, it is efficient to combine the loss functions at utterance level while keeping the attention for synchronizing between complete utterances and their subsequences. However, such a synchronization implementation is usually less memory efficient or has a largely increased training time. We experimentally found that training a LSTM model with CE loss on entire utterances also gives performance comparable to the use of subsequences (even though the training time increases due to less parallelization optimization). So we propose to compute and combine CE loss and CTC loss over entire utterances. The combined loss function is explicitly defined by involving a tunable parameter $\lambda$ as following:

\vspace{-0.4cm}
\begin{eqnarray*}
	\mathcal{L}_{MLT} = (1 - \lambda)\mathcal{L}_{CTC} + \lambda \mathcal{L}_{CE} \\ 
\end{eqnarray*}
\vspace{-1.2cm}

\vspace{-0.2cm}
\subsection{Training Optimizations}
The optimizations for CTC training was early found to be challenging in which the convergence of CTC models are unstable and largely depended on good initializations \cite{sak2015learning}. \cite{sak2015learning,sak2015fast} used parameters initialized from a framewise CE model to stabilize their CTC training. Several works \cite{amodei2016deep,miao2015eesen,audhkhasi2018building,yu2018multistage} have adopted a curriculum learning strategy in which training utterances are sorted ascending by frame length to improve the stability and the accuracy of model training. In \cite{audhkhasi2017direct}, the authors have shown that curriculum learning is not enough to train CTC model using natural words as label units. On the same work, they have presented that model initialization through pretraining is critical and a random initialization of model parameters usually fails to converge due to the occurrence of data sparsity problem. \cite{audhkhasi2017direct,audhkhasi2018building,yu2018multistage} have used a pretrained phone model and GloVe word embeddings \cite{pennington2014glove} to initialize their CTC word models.

So far, the earlier studies have used the framewise CE to initialize a part of CTC model \cite{sak2015learning,sak2015fast} or used it as an auxiliary task to accelerate the main CTC task in a jointly training \cite{yu2018multistage}. In this study, we revisit the multi-task learning of CTC and framewise CE criteria, find the optimizations for learning shared representation while considering both tasks can be the main task. Using the multi-task learning structure proposed in Section \ref{ssec:join_learning}, we experimented with the CTC tasks on vary label sets (phoneme and grapheme and word units) and the conventional framewise CE task of classifying context-dependence phones. Our training optimization includes two steps. First, we seek for the optimal value of $\lambda$ and train the multi-task models until the combined loss converges. We then perform fine-tuning on the individual tasks to have the final models. In addition to two-step training, we also evaluated several combinations of training optimizations for the multi-task learning.

\section{Experimental Setup}
\label{sec:setups}
Our experiments were conducted on the Switchboard-1 Release 2 training corpus which contains 300 hours of speech. The Hub5'00 evaluation data was used as test set. We used a deep bidirectional LSTM with 5 layers of 320 units (big models with 500 units). All the models were trained on 40 log mel filter-bank features which are normalized per conversation.
In multi-task, plain CTC and framewise CE training, we used stochastic gradient descent (SGD) while for attention-based model, we used Adam \cite{kingma2014adam}. In addition to that, new-bob training schedules are adopted to optimize all our training models. Initial learning rates are set as high as possible for individual models. For SGD, we use a decaying factor of 0.8, while it is 0.25 for Adam. 
We used the PyTorch toolkit \cite{paszke2017automatic} to build and train the deep LSTM models. The character and phoneme CTC systems were decoded using Eesen \cite{miao2015eesen} while for the hybrid systems, we used Janus \cite{Janus}. A 4-gram language model is used in all decodings (except for CTC word and attention-based models), which is trained on the transcripts of the training dataset and English Fisher corpus.

To reduce the length of input sequences in attention-based model training, we adopted down-sampling on acoustic features together with a max-pooling of two time-steps before the highest layer of LSTM models. The down-sampling is performed by stacking two consecutive frames followed by the drop of one frame. We used the attention mechanism described in \cite{bahdanau2014neural}, and a fixed rate of 0.3 when implementing Sampling \cite{bengio2015scheduled}. The beam size used in decoding is set to 12.

\section{Results}
\label{sec:results}

\subsection{Optimal Value of $\lambda$}
\label{ssec:baseline_ctc}

We observed that when $\lambda$ is set to about $[0.9 \pm 0.05]$, both token error rate (TER) and phone error rate (PER) measured on the CTC and framewise CE tasks decrease faster than in the training of the plain models. This value of $\lambda$ is stable for different sets of CTC label units and different sizes of context-dependence phones. This observation indicates an optimal value of $\lambda$ in which the learning of the shared network can maximally benefit.

\subsection{Performance of CTC Task}
\label{ssec:results_ctc}

In Table \ref{tab:ctc_baseline}, we summarize the results from recent studies which reported their CTC systems and training optimizations on the Switchboard 300 hours training set. So far, the effective training optimization typically includes the usage of a pretrained model or the order selection of training utterances. For fair comparisons, the selected phoneme and grapheme systems use 45 English phones and 46 characters as the label unit sets while the decoding was performed with n-gram language models. The word models (A2W) were trained with 10k label units and only greedy decoding is applied. Note that the presented systems employ slightly different network models and input features. \cite{amodei2016deep} and \cite{yu2018multistage} added two and three CNN layers in front of the LSTM layers, while \cite{audhkhasi2017direct, audhkhasi2018building} used i-vectors and deltas as additional input features.

\begin{table}[t]
	\caption{Baseline systems trained with CTC task using phoneme, grapheme and word labels on Hub5'00 test set.}
	\label{tab:ctc_baseline}
	\vspace{-0.2cm}
	\setlength{\tabcolsep}{3pt}
	\centering
	\begin{tabular}{lccccc}
		\toprule
		\textbf{Model} & \textbf{Pre-training} & \textbf{Train. Order} & \textbf{SWB} & \textbf{CH} \\
		\midrule
		Phone \cite{miao2015eesen} & \textit{N} & Ascending & 14.1 & 25.7 \\
		Phone \cite{audhkhasi2017direct} & \textit{N} & Descending & 14.5 & 25.1 \\
		\midrule 
		Char \cite{miao2015eesen} & \textit{N} & Ascending & 17.3 & 31.0 \\
		Char \cite{amodei2016deep} & \textit{N} & Random & 20.0 & 31.8 \\
		\midrule
		A2W \cite{audhkhasi2017direct} & Phone+GloVe & Descending & 20.8 & 30.4 \\
		A2W \cite{audhkhasi2018building} & Phone+GloVe & Ascending & 14.9 & \textbf{23.8} \\
		A2W \cite{yu2018multistage} & Phone+GloVe & Ascending & 14.8 & 25.8 \\
		A2W+CE \cite{yu2018multistage} & Phone+GloVe & Ascending & \textbf{14.3} & 25.0 \\		
		\bottomrule
	\end{tabular}
	\vspace{-0.4cm}
\end{table}

Table \ref{tab:results_ctc} presents the results of the CTC systems optimized with the proposed multi-task learning. We also provide the results of our plain CTC training with the same label sets. The performance of our plain models on phonemes and characters are at par with \cite{miao2015eesen}. We tried to randomly initialize word models, but it was not effective. However, when using the LSTM layers from the pretrained phone model, the word model training converged successfully.

Using $\lambda=0.9$, we trained and evaluated many multi-task models with different training optimizations. When the common curriculum learning strategy is applied, we observe the performance of the CTC task of the multi-task models does not significantly improve over the plain models. However, randomly shuffle training utterances together with employing parameter dropout \cite{hinton2012improving} of LSTM layers, the models generalize significantly better. With this optimization, we achieved a 13.2\% rel. WER improvement on the SWB sub-set for both character and word models compared to the plain models. This can be explained as curriculum learning is usually effective due to the training of CTC models is not stable (i.e. validation loss can change dramatically between two consecutive epochs). For the multi-task learning, the framewise CE task plays the role of stabilizing the training, so that the CTC task converges in more stable manner, and both tasks concurrently learn better generalization with the proposed optimization.

We also found that learning a word model jointly with framewise CE solved the problem of data sparsity. Without any pretrained initialization, our word models converged as well as phone or character models. This shows an interesting result since we can directly train an acoustic-to-word model which is different from \cite{audhkhasi2017direct,audhkhasi2018building,yu2018multistage}.

After the training of the multi-task models converged, we tried to fine-tune the individual tasks. This is equal as training plain models with pretrained parameters. As shown in Table \ref{tab:results_ctc}, plain word models were initialized with the parameters from the multi-task model extracted at different training epochs (labeled as \textit{m-pretrain-epoch}). We were able to train the plain models with the proposed optimization, although it was hard to gain further improvement. This indicates that the framewise CE not only stabilizes the training but also leads to the learning of a shared representation which is effective for the CTC criterion.

\begin{table}[t]
	\caption{The performance of our plain CTC models and multi-task CTC models on Hub5'00 test set.}
	\label{tab:results_ctc}	
	\vspace{-0.2cm}	
	\setlength{\tabcolsep}{2.5pt}
	\centering
	\begin{tabular}{lccccc}
		\toprule
		\textbf{Model} & \textbf{Train. Order} & \textbf{TER} & \textbf{PER} & \textbf{SWB} & \textbf{CH} \\
		\midrule
		Char & Ascending & 15.0 & - & 17.2 & 29.6 \\		
		Phone & Ascending & 15.1 & - & 14.5 & 25.9 \\
		A2W(\textit{p-pretrain}) & Ascending & 23.6 & - & 18.8 & 29.1 \\
		A2W(\textit{p-pretrain}) & Random & 25.1 & - & 20.8 & 31.1 \\		
		\midrule
		Char+CE & Ascending & 15.0 & 29.8 & 16.8 & 29.3 \\
		Phone+CE & Ascending & 14.3 & 29.6 & 13.8 & 25.3 \\
		A2W+CE & Ascending & 22.5 & 29.2 & 18.0 & 28.8 \\
		Char+CE & Random & 14.1 & 28.3 & \textbf{15.1} & \textbf{27.2} \\
		Phone+CE & Random & 13.5 & 27.5 & \textbf{13.4} & \textbf{24.5} \\		
		A2W+CE & Random & 20.4 & 26.7 & 16.5 & 27.0 \\
		A2W+CE(big) & Random & 19.8 & 25.8 & \textbf{15.8} & \textbf{26.2} \\				
		\midrule
		A2W(\textit{m-pretrain-3}) & Random & 22.6 & - & 17.9 & 29.0 \\
		A2W(\textit{m-pretrain-10}) & Random & 20.5 & - & 16.6 & 27.0 \\		
		A2W(\textit{m-pretrain-20}) & Random & 20.3 & - & 16.3 & 26.9 \\
		\bottomrule
	\end{tabular}
	\vspace{-0.3cm}
\end{table}

\begin{table}[t]
	\caption{The performance of hybrid systems with and without multi-task training. All the models use dropout regularization.}
	\label{tab:results_ce}
	\vspace{-0.2cm}	
	\setlength{\tabcolsep}{3pt}
	\centering
	\begin{tabular}{lccccc}
		\toprule
		\textbf{Model} & \textbf{Sub-seq.} & \textbf{Train. Order} & \textbf{PER} & \textbf{SWB} & \textbf{CH} \\
		\midrule
		FFNN \cite{zeyer2017comprehensive} & Y & Random & - & 13.1 & 25.6 \\
		LSTM(big) \cite{zeyer2017comprehensive} & Y & Random & - & \textbf{11.9} & \textbf{22.3} \\		
		\midrule
		FFNN & Y & Random & 54.9 & 15.7 & 27.9 \\
		LSTM(big) & Y & Random & 28.3 & 13.6 & 23.8 \\
		LSTM & N & Ascending & 31.0 & 14.3 & 24.9 \\
		LSTM & N & Random & 28.4 & 13.8 & 24.4 \\
		A2W+CE & N & Random & 26.7 & 12.8 & 22.9 \\
		A2W+CE(big) & N & Random & 25.8 & \textbf{12.0} & \textbf{22.4} \\
		\bottomrule
	\end{tabular}
	\vspace{-0.6cm}
\end{table}

\subsection{Performance of Framewise CE Task}

As earlier mentioned, we are also interested in optimizing and evaluating the framewise CE task in the multi-task training. From the phone error rates shown in Table \ref{tab:results_ctc}, it is interesting to see that the framewise CE task achieves better performance when training together with the word-based CTC task. Taking this model, we compared it with several LSTM hybrid acoustic models as presented in Table \ref{tab:results_ce}. The presented models were trained either on entire utterances or subsequences. Our setup for constructing subsequences is similar to the optimal setup found in \cite{zeyer2017comprehensive}, in which training utterances are divided into chunks of 50 frames while two consecutive chunks have 25 overlapping frames.
In term of WER, we achieved a significant improvement (12\% rel.) between the multi-task training and the plain training. The result of our multi-task model with bigger size is at par with the best model reported in \cite{zeyer2017comprehensive}, while this model employed more advanced input features.

Our finding in using word-based CTC task to supplement and boost the performance of the hybrid model consolidates the effectiveness of learning shared representation for both tasks. To best of our knowledge, we are not aware of earlier works which have reported the result.

\vspace{-0.2cm}
\section{Pre-trained Encoder for Attention-based Model}

\begin{table}[t]
	\caption{The performance of attention-based models with and without pretrained encoder.}
	\label{tab:results_attn}
	\vspace{-0.2cm}	
	\setlength{\tabcolsep}{3pt}
	\centering
	\begin{tabular}{lccc}
		\toprule
		\textbf{Model} & \textbf{Unit} & \textbf{SWB} & \textbf{CH} \\
		\midrule
		3Enc-1Dec \cite{lu2016training} & Char & 27.3 & 48.2 \\				
		3Enc-1Dec \cite{lu2016training} & Word & 26.3 & 46.5 \\		
		4Enc-1Dec \cite{toshniwal2017multitask} & Char & 23.1 & 40.8 \\	
		6Enc-1Dec \cite{zeyer2018improved} & BPE & 13.1 & 26.1 \\
		6Enc-1Dec (+LSTM LM) \cite{zeyer2018improved} & BPE & \textbf{11.8} & 25.7 \\
		6Enc-2Dec (+Speed perturbation) \cite{weng2018improving} & Char & 12.2 & \textbf{23.3} \\		
		\midrule
		6Enc-1Dec (-without Pre-training) & Char & 16.9 & 28.9 \\
		6Enc-1Dec (-without Pre-training)& Word & 18.8 & 30.4 \\
		6Enc-1Dec & Char & 16.3 & 27.9 \\
		6Enc-1Dec & Word & 14.7 & 26.1 \\
		8Enc-2Dec & Word & 14.1 & 25.5 \\
		10Enc-2Dec(big) & Word & 13.6 & 24.9 \\
		\;\; +Speed perturbation & Word & 12.5 & 23.1 \\
		\;\;\;\; +Sampling & Word & 12.2 & \textbf{22.6} \\
		\bottomrule
	\end{tabular}
	\vspace{-0.4cm}
\end{table}

We have presented that useful representation can be shared for many speech recognition models built with CTC and framewise CE training criteria. This shows a fact that different neural network models optimized to perform the speech recognition task may potentially learn similar traits in representation. In this section, we investigate the usage of the representation learned by our proposed multi-task model to train attention-based models.

Attention-based speech recognition model \cite{chorowski2015attention,bahdanau2016end,chan2016listen} is a single neural network that consists of an encoder RNN and a decoder RNN, and uses attention mechanism to connect between them. The decoder is analogous to a language model due to attention-based model is trained to provide a probability distribution over sequences of labels (words or characters). The encoder converting low level acoustic features into higher level representation is analogous to the RNN of CTC model. In fact, the encoder is supposed to extract features not only for the prediction of label sequence but also for the alignments provided by the attention function. In this sense, the encoder needs to learn complex representation so that the training of attention-based model usually converges slower than CTC or framewise hybrid models. To improve the convergence of attention-based model, we propose to initialize the encoder with the parameters from the pretrained multi-task model (with word-based CTC and CE tasks). Basically we take the whole shared LSTM part and the projection layer, and then stack some more LSTM layers on top to form the encoder of our attention-based models.

\begin{figure}[t]
	\centering
	\includegraphics[width=\linewidth]{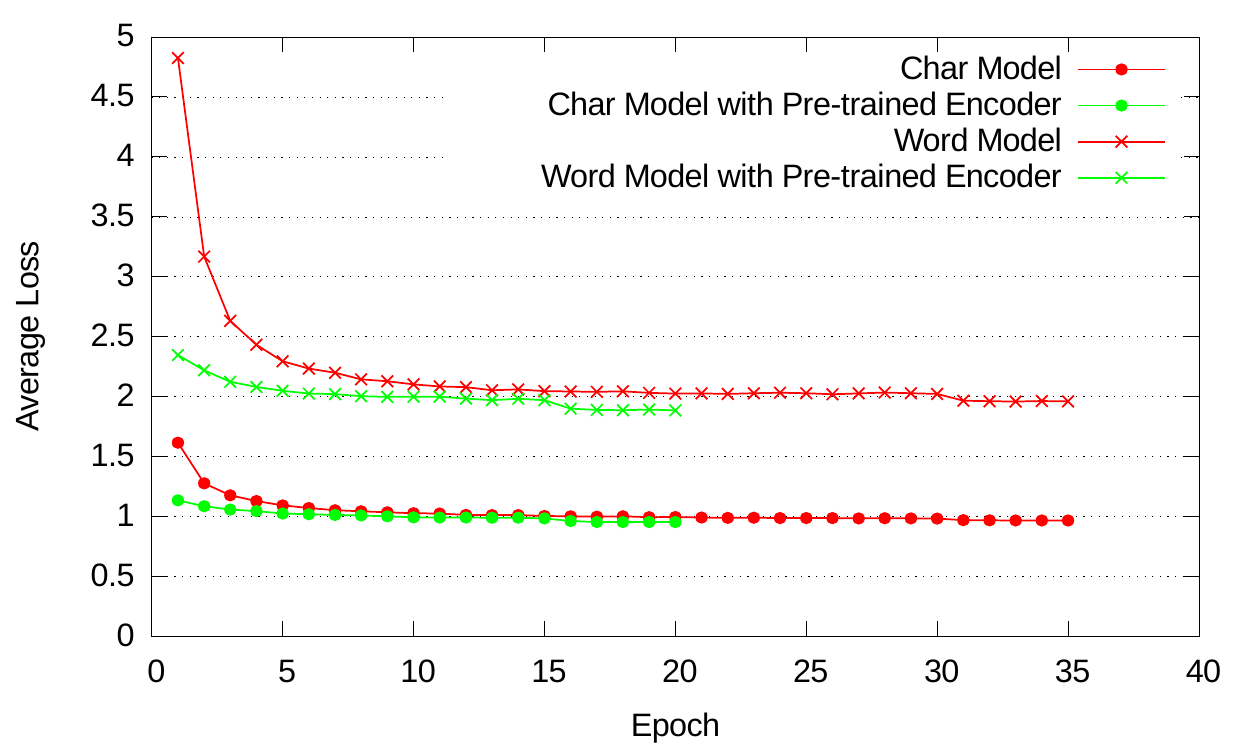}
	\vspace*{-0.5cm}	
	\caption{{Cross-validation loss over training epochs.}}
	\label{fig:att_loss}
	\vspace*{-0.6cm}
\end{figure}

We have trained many attention-based models with and without pretraining initialization using either word or character labels. In our experiments, the regular models (with random initializations of parameters) were trained for 35 epochs while for the pretrained models, we used only 20 epochs (see Section \ref{sec:setups} for detailed optimizations). Figure \ref{fig:att_loss} shows the convergence of the training models with the average loss over training epochs. As can be seen, the models with pretraining converge much faster than without (e.g., to reach to the same state the pretrained models need 3 epochs while it is 20 epochs for regular models). In our setup, the pretraining initialization also make the models eventually converge at a better final state. 

The other advantage of using pretrained layers is that we can stack several LSTM layers while keeping the same convergence manner. As shown in Table \ref{tab:results_attn}, we could achieve further improvement when train an acoustic-to-word attention-based model with a deep encoder of 10 layers. When employing Speed perturbation~\cite{ko2015audio} and Sampling~\cite{bengio2015scheduled}, our deep model generalizes much better, which then outperforms the previous works reporting on CallHome subset of the Hub5'00. We did not try with other techniques proposed in \cite{weng2018improving}.

\vspace{-0.2cm}
\section{Related Work}
\label{sec:related_work}

\cite{sak2015learning,sak2015fast} used framewise CE training to initialize the LSTM layers of phone-based CTC models. They found that using such pretrained parameters, CTC training is more stable than when using random initialization. \cite{zhang2018deep} trained deep feed-forward sequential memory networks (Deep-FSMN) with CTC and proposed to incorporate CE loss as a regularization term. They argued that CE loss is helpful in stabilizing CTC training and improving the alignments of CTC models, which then lead to significant improvements in WER. 

Current work \cite{yu2018multistage} proposed to use hierarchical pretrained CTC and curriculum learning and a joint CTC-CE training to optimize A2W models. In \cite{yu2018multistage}, multi-task CTC-CE was also investigated but did not produce performance gain. In our study, we found that the proposed training optimization is needed for improving the convergence of multi-task models and the usage of pretraining is not critical for training acoustic-to-word.

\cite{hori2017advances} presented that CTC and attention-based models can share encoder's representation. We additionally showed pretraining an encoder also helps attention-based model converge faster and better. 

\vspace{-0.2cm}
\section{Conclusion}
\label{sec:conclusion}

We have presented an efficient approach for training encoder networks for modeling CTC and conventional framewise CE criteria. We further showed that end-to-end speech recognition such as acoustic-to-word and attention-based systems can benefit from sharing encoding representation.

\bibliographystyle{IEEEtran}

\bibliography{mybib}

\end{document}